\documentclass[a4paper,11pt]{article}
\usepackage{jcappub}
\usepackage{amsmath}
\usepackage{graphicx}
\usepackage{latexsym}
\usepackage{xspace}
\usepackage{color}
\usepackage{bm}
\usepackage{relsize}
\usepackage{tabularx}
\usepackage{multirow}
\usepackage{amssymb}
\usepackage[table]{xcolor}
\usepackage{braket}
\usepackage{comment}
\usepackage[utf8]{inputenc}
\usepackage{slashed}
\usepackage{empheq}
\usepackage{subfigure}
\usepackage{here}

\usepackage[ruled,vlined]{algorithm2e}

\makeatletter
\gdef\@fpheader{}
\g@addto@macro\bfseries{\boldmath}
\makeatother

\newcommand{\prn}[1]{\left( {#1} \right)}
\newcommand{\com}[1]{\left[ {#1} \right]}

\newcommand{\abs}[1]{\left\vert {#1} \right\vert}

\newcommand{\dif}[2]{\frac{\mathrm{d} #1}{\mathrm{d} #2}}
\newcommand{\pdif}[2]{\frac{\partial #1}{\partial #2}}

\newcommand{\uir}{\mathrm{IR}}
\newcommand{\uuv}{\mathrm{UV}}

\newcommand{\udet}{{}_\mathrm{det}}
\newcommand{\usto}{{}_\mathrm{sto}}

\newcommand{\ie}{\textsl{i.e.~}}

\newcommand{\dd}{\mathrm{d}}

\newcommand{\sss}[1]{{\scriptscriptstyle{#1}}}

\newcommand{\uPl}{\mathrm{Pl}}

\newcommand{\usssPl}{\sss{\uPl}}

\newcommand{\mpl}{M_\usssPl}

\newcommand{\efolds}{$e$-folds}

\newcommand{\beq}{\begin{equation}}
\newcommand{\eeq}{\end{equation}}
\newcommand{\bea}{\begin{equation}\begin{aligned}}
\newcommand{\eea}{\end{aligned}\end{equation}}

\newlength{\wsingfig}
\newlength{\wdblefig}
\newlength{\wquadfig}
\newlength{\wtriplefig}
\newcommand{\Eq}[1]{Eq.~(\ref{#1})}
\newcommand{\Eqs}[1]{Eqs.~(\ref{#1})}

\newcommand{\Sec}[1]{Sec.~\ref{#1}}

\newcommand{\App}[1]{Appendix~\ref{#1}}

\subheader{}

\title{A consistent formulation of stochastic inflation I: Non-Markovian effects and issues beyond linear perturbations}

\author[a]{Diego Cruces,}
\author[b,c]{Tomotaka Kuroda}

\affiliation[a]{Institute of Theoretical Physics, Chinese Academy of Sciences, Beijing 100190, China}
\affiliation[b]{Department of Physics, Institute of Science Tokyo Tokyo, 152-8551, Japan}
\affiliation[c]{Cosmology, Gravity and Astroparticle Physics Group, Center for Theoretical Physics of the Universe, Institute for Basic Sclence (IBS), Daejeon, 34126, Korea}

\emailAdd{dcruces@itp.ac.cn}
\emailAdd{kuroda.t.ad@m.titech.ac.jp}

\date{today}

\begin{document}
\sloppy

\abstract{We investigate the origin of non-Markovianity in stochastic inflation and its implications for nonlinear perturbation theory. In the Schwinger--Keldysh formulation, the noise terms sourcing the infrared (IR) Langevin equations are determined by ultraviolet (UV) modes evolving on top of the stochastic IR background. Since the UV-mode evolution generally depends on the past history of the IR sector, the resulting stochastic dynamics is intrinsically non-Markovian. Working perturbatively, we derive the UV-mode solutions up to second order and decompose the corresponding noise contributions into two parts. The first is a ``deterministic'' contribution, generated by the functional Taylor expansion of the first-order UV solution around the background trajectory. The second is a genuinely ``stochastic'' contribution, originating from terms in the UV-mode equations that are quadratic in the noise variables and are usually neglected in the standard formulation of stochastic inflation. Under this conventional truncation, the deterministic contribution reduces to a Markovian correction in attractor backgrounds, whereas it could become history dependent in non-attractor phases and gives rise to non-Markovian terms involving integrals over first-order IR perturbations. We finally show that the stochastic contribution is of the same perturbative order as the deterministic one, which indicate that the conventional truncation is generically inconsistent and quadratic-noise terms may be required for a consistent treatment of nonlinear perturbations in stochastic inflation. Our analysis clarifies the perturbative structure of non-Markovianity and provides the basis for a systematic treatment of quadratic-noise effects beyond the standard formulation.}

\arxivnumber{}

\maketitle


\flushbottom


\section{Introduction}\label{Sec:1}

Cosmological inflation is a period of accelerated expansion in the very early Universe. Quantum fluctuations generated during inflation in an approximately de Sitter spacetime provide a compelling explanation for a variety of observed phenomena, most notably the temperature anisotropies of the cosmic microwave background (CMB). A precise understanding of the quantum dynamics of quasi-de Sitter spacetime, including radiative corrections, is therefore essential for deriving reliable first-principles predictions for cosmological observables.

In this paper, we employ the stochastic formalism~\cite{Starobinsky:1986fx,Starobinsky:1994bd} as the theoretical framework for describing quantum perturbations in this setting. The stochastic formalism can be regarded as an effective field theory for infrared (IR) degrees of freedom, in which the effects of ultraviolet (UV) fluctuations are systematically incorporated through stochastic noise~\cite{PerreaultLevasseur:2013kfq,Burgess:2014eoa,Burgess:2015ajz,Moss:2016uix,Grain:2017dqa,Gorbenko:2019rza,Baumgart:2019clc,Mirbabayi:2019qtx,Mirbabayi:2020vyt,Cohen:2020php,Cohen:2021fzf,Cruces:2018cvq,Cruces:2021iwq,Cruces:2022imf,Li:2025azq} . This formalism makes it particularly well suited for treating the infrared divergences that arise for light spectator fields in de Sitter spacetime, which constitute a well-known limitation of conventional perturbative methods. Within the stochastic approach, the relevant IR loop contributions are effectively resummed, and the secular growth appearing in perturbation theory is reinterpreted as the growth of statistical variances~\cite{Tsamis:2005hd,Enqvist:2008kt,Seery:2010kh,Garbrecht:2013coa,Garbrecht:2014dca,Tokuda:2017fdh,Tokuda:2018eqs,Honda:2023unh,Cespedes:2023aal,Palma:2023idj,Palma:2023uwo,Palma:2025oux}. Moreover, the formalism allows one to compute probability distribution functions (PDFs) directly. Since PDFs encode the information contained in correlation functions of arbitrarily high order~\cite{Nakao:1988yi,Finelli:2008zg,Markkanen:2019kpv,Markkanen:2020bfc,Cable:2020dke,Cable:2022uwd,Cable:2023gdz,Bhattacharya:2022wjl}, this provides a substantial technical advantage over perturbative approaches, which typically require a truncation at finite order.

The stochastic formalism has also recently found important applications in the study of cosmological perturbations in nonlinear, and even genuinely non-perturbative, regimes. This is particularly relevant for phenomena such as primordial black holes (PBHs) and scalar-induced gravitational waves, which are sourced by large-amplitude fluctuations and therefore require a description beyond conventional perturbation theory. In this context, the stochastic-$\delta N$ formalism~\cite{Fujita:2013cna,Fujita:2014tja,Vennin:2015hra} provides a non-perturbative framework for characterizing curvature perturbations in terms of fluctuations in the local amount of inflation. These fluctuations are quantified by the number of inflationary~\efolds, through the geometrical $\delta N$ relation~\cite{Starobinsky:1982ee,Starobinsky:1986fxa,Sasaki1996,Sasaki:1998ug,Lyth:2004gb}. From a technical perspective, the statistics of the number of \efolds\ can be formulated as a first-passage-time problem~\cite{Vennin:2015hra,Pattison:2017mbe,Ezquiaga:2019ftu}. This formulation naturally gives rise to strongly non-Gaussian probability distributions, often with pronounced heavy tails. Such heavy-tailed statistics are especially important in the context of PBH formation, since they can significantly enhance the probability that curvature perturbations exceed the collapse threshold. For this reason, the stochastic-$\delta N$ formalism has been applied to a broad class of inflationary models relevant to PBH production~\cite{Ballesteros:2020sre,Figueroa:2020jkf,Figueroa:2021zah,Tomberg:2023kli,Mishra:2023lhe,Tada:2023fvd,Mizuguchi:2024kbl,Murata:2025onc,Kuroda:2025coa,Raatikainen:2025gpd}. The corresponding tail probabilities can be evaluated using a variety of analytical and numerical techniques~\cite{Ando:2020fjm,Tada:2021zzj,Kitajima:2021fpq,Hooshangi:2021ubn,Gow:2022jfb,Raatikainen:2023bzk,Animali:2024jiz,Vennin:2024yzl,Jackson:2024aoo,Animali:2025pyf,Choudhury:2025kxg,Murata:2026yqb,Saha:2026cay}. It should be emphasized, however, that most existing analyses, including those based on the first-passage-time formulation, rely on the assumption that the stochastic dynamics is Markovian. As we discuss below, this assumption is not guaranteed in general, since non-Markovian effects can arise naturally in stochastic descriptions of inflationary dynamics.

A particularly systematic formulation of the stochastic formalism is based on the Schwinger--Keldysh path-integral description of open quantum systems~\cite{Morikawa:1989xz,PerreaultLevasseur:2013kfq,Moss:2016uix,Pinol:2020cdp,Panda:2025tpu}. In this approach, the UV degrees of freedom are integrated out and treated as an environment, whose effect on the IR sector is encoded in stochastic diffusion across the coarse-graining scale. The resulting IR dynamics are described by Langevin equations supplemented by statistical noise terms. These noise terms are determined by the evolution of UV modes on top of the stochastic IR background. This requires solving not only the Langevin equations for the IR variables, but also the mode equations for the UV fluctuations at all relevant scales and at each time step~\cite{Figueroa:2020jkf,Figueroa:2021zah,Launay:2025kef,Launay:2025lnc,Kawasaki:2026hnx}. This coupling between the IR stochastic dynamics and the UV mode evolution can make the resulting system nonlocal in time. Consequently, such stochastic differential equations may exhibit non-Markovian behavior~\cite{Cruces:2022imf,Figueroa:2021zah,Pinol:2020cdp,Tomberg:2022mkt,Tomberg:2024evi,Brahma:2024ycc,Calderon-Figueroa:2025dto,Kawasaki:2026hnx}, since the evolution of the IR variables can depend on the prior history of the system, rather than solely on their instantaneous state. We discuss this point in more detail in \Sec{Sec:2}.

In this work, we analytically investigate the origin and implications of non-Markovianity in stochastic inflation. Our analysis further indicates that the omission of terms quadratic in the noise variables should be reconsidered in order to obtain a consistent description of nonlinear perturbations. In order to show these, \Sec{Sec:2} is devoted to reviewing the stochastic formalism under the standard assumption that terms quadratic in the noise variables can be neglected. This formulation makes explicit the non-Markovian contribution, which becomes relevant in non-attractor models. In \Sec{Sec:3}, we then examine the validity of neglecting such quadratic-noise terms within the perturbative regime. To this end, we perturb all equations governing the stochastic system around the homogeneous background. This includes both the Langevin equations for the IR degrees of freedom and the UV-mode equations that determine the statistical properties of the noise. Within this framework, we classify the UV modes, and hence the associated noise terms, at second order. One class corresponds to a ``deterministic'' contribution, which follows directly from the functional Taylor expansion of the stochastic system introduced in \Sec{Sec:2}. The other class corresponds to a genuinely ``stochastic'' contribution, which is precisely associated with the quadratic-noise terms neglected in the conventional treatment. Finally, in \Sec{Sec:4}, we compare these two classes of contributions and show that they are generically of the same order. This observation calls into question the standard truncation in which quadratic-noise terms are discarded, and provides the motivation for a forthcoming companion paper, where we will analyze in detail the full set of such terms that are commonly overlooked in the literature.

\section{Brief review of the stochastic inflation}
\label{Sec:2}

In this section, we briefly review the stochastic formalism, paying particular attention to its detailed structure, especially the origin of non-Markovianity. This feature is distinctive to the stochastic formalism and differs from the type of non-Markovianity that arises in conventional open-system approaches.

We consider a system consisting of a single scalar field $\phi$, whose action is
\begin{align}
S=\int \dd^4 x \sqrt{-g}\com{\frac{\mpl^2}{2}R-\frac{1}{2}g^{\mu\nu}\partial_\mu\phi\partial_\nu\phi - V(\phi)} ,
\end{align}
Here, $g^{\mu\nu}$ denotes the (inverse) metric and $V(\phi)$ is the scalar potential. For later convenience, we use the number of~\efolds\ $N$, defined by $\dd N=H\dd t$, as the time variable throughout this paper.

\subsection{Formulation}
In the stochastic formalism, fields are decomposed into IR and UV modes. The separation is implemented by means of a window function $W(x)$, which at leading order is taken to be a step function, $W(x)\rightarrow\theta(1-x)$. For a generic quantity $Q$, the IR and UV components are defined as
\begin{align}
    Q_\mathrm{IR} &\equiv \int\frac{\dd^3 k}{(2\pi)^3} e^{i \mathbf{k}\cdot\mathbf{x}} W\left(\frac{k}{k_\sigma(N)}\right) Q(N,\mathbf k) \equiv \int\frac{\dd^3 k}{(2\pi)^3} e^{i \mathbf{k}\cdot\mathbf{x}} Q_\mathrm{IR}(N,\mathbf k),\\
    Q_\mathrm{UV} &\equiv \int\frac{\dd^3 k}{(2\pi)^3} e^{i \mathbf{k}\cdot\mathbf{x}} \com{1-W\left(\frac{k}{k_\sigma(N)}\right)} Q(N,\mathbf k) \equiv \int\frac{\dd^3 k}{(2\pi)^3}e^{i \mathbf{k}\cdot\mathbf{x}} Q_\mathrm{UV}(N,\mathbf k).
\end{align}
Here, the coarse-graining scale or the cutoff scale is defined by
\begin{align}
    k_\sigma(N) = \sigma a H_\mathrm{IR},
\end{align}
with $\sigma\ll 1$~\cite{Grain:2017dqa,Andersen:2021lii}. 

Intuitively, the stochastic formalism proceeds as follows. We typically work in the flat gauge, which is equivalent to the uniform-$N$ gauge on super-horizon scales under the separate-universe assumption~\cite{Sasaki:1998ug,Wands:2000dp,Lyth:2003im,Lyth:2004gb} \footnote{Throughout this paper, we adopt the definition of the separate-universe approximation used in Ref.~\cite{Cruces:2022imf}, namely the approximation in which both non-local terms and the momentum constraint of general relativity are neglected. As shown in Ref.~\cite{Cruces:2022imf}, when these contributions are retained, the spatially flat gauge and the uniform-$N$ gauge are no longer equivalent on superhorizon scales, although their difference is exponentially suppressed in most situations. A direct comparison between these two gauges on superhorizon scales, in regimes relevant to primordial black hole formation, can be found in Appendix~C of Ref.~\cite{Figueroa:2021zah}.} The UV modes are treated quantum mechanically, typically within perturbation theory and under the assumption of the Bunch--Davies vacuum. By contrast, the IR dynamics are effectively classical, and spatial gradient terms can be neglected. As a result, each coarse-grained patch evolves independently within the separate-universe approximation. As UV modes cross the coarse-graining scale $k_\sigma(N)$, they are transferred into the IR sector, thereby inducing stochastic fluctuations that drive the evolution of the IR background away from the homogeneous background. This continuous inflow of modes is modeled by stochastic noise terms, leading to Langevin equations that govern the dynamics of the IR modes.

This physical picture can be made precise within the path-integral formulation of open quantum systems~\cite{Morikawa:1989xz,PerreaultLevasseur:2013kfq,Moss:2016uix,Pinol:2020cdp,Panda:2025tpu}, although alternative derivations based on wave functionals or density matrices have also been developed~\cite{Burgess:2014eoa,Burgess:2015ajz,Moss:2016uix,Gorbenko:2019rza,Cohen:2020php,Cohen:2021fzf,Cespedes:2023aal,Calzetta:2008iqa}. Starting from the action of the full theory, one separates the degrees of freedom into an IR sector, which constitutes the system of interest, and UV-dependent terms, which encode both the environmental degrees of freedom and their interactions with the IR system. Integrating out the UV environmental modes then yields the influence action, which in general contains both real and imaginary parts%
\footnote{In stochastic inflation, however, the influence action is purely imaginary. This property arises because appropriate boundary conditions must be imposed when integrating out the UV modes~\cite{Tokuda:2017fdh,Tokuda:2018eqs}. It is a distinctive feature of stochastic inflation, originating from the fact that the same field degrees of freedom are split into the system, namely the IR modes, and the environment, namely the UV modes, by means of a cutoff, so that the transfer of modes from the UV sector to the IR sector is unidirectional.}. 
As a result, the effective dynamics of the IR sector is no longer unitary. The imaginary part of the influence action can be rewritten by introducing auxiliary fields through a Hubbard--Stratonovich transformation. These auxiliary fields are naturally interpreted as stochastic noise sources, and the resulting effective action describes the IR modes subject to stochastic forces. Taking the classical equations of motion derived from this effective action then leads to Langevin equations for the IR fields%
\footnote{In the Schwinger--Keldysh formalism, the classical fields are defined as the average of the field configurations on the forward, conventionally denoted by ``$+$'', and backward, conventionally denoted by ``$-$'', time contours. These are also referred to as retarded fields.}, given by
\begin{align}
    \dif{\phi_\uir}{N} &= {\pi_\uir} + \Xi_\phi,  \label{phi IR}\\
    \dif{\pi_\uir}{N} &= -\left(3-\frac{{\pi_\uir}^2}{2\mpl^2}\right)\left(\pi_\uir+\mpl^2\frac{V_{,\phi}(\phi_\uir)}{V(\phi_\uir)}\right) + \Xi_\pi,\label{pi IR}
\end{align}
where $V_{,\phi}(\phi_{\mathrm{IR}})$ denotes the derivative of the potential with respect to $\phi_\uir$. Note that, in our notation, the field and momentum variables have the same dimension. The Hubble parameter is determined by the Friedmann equation:
\begin{align}
    3\mpl^2 H_\uir^2 = \frac{H_\uir^2\pi_\uir^2}{2} + V(\phi_\uir).
\end{align}
The auxiliary fields $\Xi_{\phi}$ and $\Xi_{\pi}$ act as stochastic noise terms. At leading order, these noise terms obey Gaussian statistics, with variance\footnote{For simplicity, we focus on the dynamics of a single coarse-grained patch by invoking the separate-universe approximation. Consequently, the sinc function associated with correlations between patches at different spatial locations is neglected in the noise variance.} given by
\begin{align}
    \Braket{\Xi_{X}(N)\Xi_{Y}(N')} &= \dif{\ln k_\sigma}{N}\mathcal{P}_{XY}(N,k_\sigma(N))\delta(N-N')\notag\\
    &={\left.\prn{\prn{1-\epsilon_{1,\uir}}\frac{k^3}{2\pi^2}  (X_\mathrm{UV}Y^{*}_\mathrm{UV})}\right|_{k=k_\sigma(N)}} \delta(N-N'), \label{variance of noise}
\end{align}
where $X,Y\in\{\phi,\pi\}$, and the slow-roll parameter is defined by
\begin{align}
    \epsilon_{1,\uir} \equiv - \dif{\ln H_\uir}{N}.
\end{align}
Equivalently, one may define normalized noise variables $\xi_{X}$, which are white noises owing to the use of a step function as the window function:
\begin{align}
    \Xi_{X}(N) = \mathsf{N}_X\xi_{X}(N)\quad \mathrm{with}\ \Braket{\xi_{X}(N)\xi_{X}(N')}=\delta(N-N'), \label{normalized noises}
\end{align}
where
\begin{align}
    \mathsf{N}_\phi \equiv \sqrt{\left.\prn{\prn{1-\epsilon_{1,\uir}}\frac{k^3}{2\pi^2}\abs{\phi_\uuv}^2}\right|_{k=k_\sigma(N)}},\quad \mathsf{N}_\pi \equiv {\sqrt{\left.\prn{\prn{1-\epsilon_{1,\uir}}\frac{k^3}{2\pi^2}\abs{\pi_\uuv}^2}\right|_{k=k_\sigma(N)}}}.
\end{align}

As emphasized above, UV modes affect the IR dynamics when they cross the coarse-graining scale $k_\sigma(N)$. Note that, in order to compute the noise terms, one needs the values of the UV modes at the coarse-graining scale. These are obtained by evolving the UV modes from deep inside the horizon to the coarse-graining scale according to their equations of motion on top of the evolving IR background. The equation of motion for UV modes could be obtained by Taylor expanding them over the IR background 
\begin{align}
     0&=\pdif{\phi_\uuv}{N} - {\pi_\uuv} ,  \label{phi UV taylor 2}\\
    0&= \pdif{\pi_\uuv}{N} +\frac{\partial}{\partial\pi_\uir}\left[\left(3-\frac{{\pi_\uir}^2}{2\mpl^2}\right)\left(\pi_\uir+\mpl^2\frac{V_{,\phi}(\phi_\uir)}{V(\phi_\uir)}\right)\right]{\pi_\uuv} \notag\\  &\hspace{15mm}+\frac{\partial}{\partial\phi_\uir}\left[\left(3-\frac{{\pi_\uir}^2}{2\mpl^2}\right)\left(\pi_\uir+\mpl^2\frac{V_{,\phi}(\phi_\uir)}{V(\phi_\uir)}\right)\right]\phi_\uuv + \frac{k^2}{a^2H_\uir^2}\phi_\uuv .\label{pi UV taylor 2}
\end{align}

The equation of motion for UV modes \Eqs{phi UV taylor 2}--\eqref{pi UV taylor 2}, supplemented by the Bunch--Davies vacuum conditions deep inside the horizon, is the one that should be solved to compute the noises \eqref{variance of noise} (actually, this is what we will do in Sections \ref{Sec:3} and \ref{Sec:4}). However, in order to make connection with previous work where the non-Markovian nature of stochastic inflation is studied (such as \cite{Figueroa:2020jkf,Figueroa:2021zah,PerreaultLevasseur:2014ziv}), we can use the stochastic equation of motion for the IR modes \eqref{phi IR}--\eqref{pi IR} to rewrite \Eqs{phi UV taylor 2}--\eqref{pi UV taylor 2} as\footnote{Note that here we are also assuming that the difference between spatially flat gauge and uniform-N gauge will always be negligible, otherwise the equation for UV modes is slightly different (see \cite{Cruces:2021iwq, Cruces:2022imf, Cruces:2024pni})}

\begin{align}
     \mathcal{O}(\Xi_i)&=\pdif{\phi_\uuv}{N} - {\pi_\uuv} ,  \label{phi UV}\\
    \mathcal{O}(\Xi_i)&= \pdif{\pi_\uuv}{N} + \left(3-\epsilon_{1,\uir}\right)\pi_\uuv + \com{\frac{V_{,\phi\phi}(\phi_\uir)}{H^2_\uir} - \frac{1}{a^3\mpl^2 H_\uir}\dif{}{N}\prn{a^3 H_\uir \pi_\uir^2}} \phi_\uuv + \frac{k^2}{a^2H_\uir^2} \phi_\uuv,
    \label{pi UV}
\end{align}
where $\mathcal{O}(\Xi_i)$ denotes terms in which the noises $\Xi_\phi$ and $\Xi_\pi$ explicitly appear. Such terms are typically neglected in the literature \cite{Figueroa:2020jkf,Figueroa:2021zah,PerreaultLevasseur:2014ziv}. The reason is that they would effectively manifest as quadratic noise in the original stochastic system in \Eqs{phi IR}–\eqref{pi IR}. Since quadratic noises have been discarded when formulating stochastic inflation, it seems reasonable to also neglect $\mathcal{O}(\Xi_i)$ in the equation of motion for UV modes. In this section, we will follow the standard formulation of stochastic inflation and study non-Markovian effects under this approximation. The consequences of neglecting quadratic noises will be examined perturbatively in Sections \ref{Sec:3} and \ref{Sec:4}.

Under the same approximation, we can rewrite \Eqs{phi UV}--\eqref{pi UV} in a more compact way:

\begin{equation}
\frac{1}{a^3 H_\uir \epsilon_{1,\uir}}\dif{}{N} \left[a^3 H_\uir \epsilon_{1,\uir}\dif{}{N}\left(\frac{\phi_\uuv}{\pi_\uir}\right)\right] + \frac{k^2}{a^2H_\uir^2}\frac{\phi_\uuv}{\pi_\uir}=\mathcal{O}(\Xi_i).
\label{phi UV compact}
\end{equation}


\subsection{The method to solve the system}
\label{sec:post:inflation}

Let us now comment on the structure of the resulting equations. The UV mode equation, \eqref{phi UV compact}, is constructed to be formally analogous to the linear equation of motion familiar from cosmological perturbation theory. The crucial difference is that the fixed, deterministic background is replaced by the stochastic IR background. Consequently, the UV mode equation cannot, in general, be solved analytically, and the corresponding UV power spectra are not available in analytical form. In order to determine the noise terms at each time step, one must solve the UV mode equation \eqref{phi UV compact} on top of the IR background evolving according to Eqs.~\eqref{phi IR}--\eqref{pi IR}. Since the solution of Eq.~\eqref{phi UV compact} depends on the full prior evolution of the IR background, the noise terms required to update the IR variables at a given time generically retain memory of the earlier history of the system. In other words, the values of the IR variables at a single instant are not sufficient to determine their evolution to the next time step through the Langevin equations \eqref{phi IR}--\eqref{pi IR}. By definition, the resulting stochastic process is therefore non-Markovian. A fully consistent treatment of this non-Markovian dynamics generally requires numerical simulations in which the UV-mode equations are solved for all relevant modes at each time step, together with the stochastic evolution of the IR variables~\cite{Figueroa:2020jkf,Figueroa:2021zah,Launay:2025kef,Launay:2025lnc,Kawasaki:2026hnx}. Such simulations are computationally demanding. There also exist practical alternatives, such as recursive methods~\cite{PerreaultLevasseur:2013eno,PerreaultLevasseur:2013kfq,Cruces:2024pni,PerreaultLevasseur:2014ziv,Figueroa:2021zah}, which can partially circumvent the need for the full numerical procedure described above.

However, if the characteristic time scale over which the UV modes evolve as the environment is much shorter than that of the IR modes as the system, then the UV dynamics at a given time is effectively sensitive only to the instantaneous values of the IR modes, This is precisely what happens at subhorizon scales $k \gg a H_\uir$, implying that any memory effects due to the subhorizon evolution of UV modes will be suppressed. On the other hand, the timescale of UV modes near the coarse-graining scale can be comparable to that of IR modes, because the noise terms are evaluated from UV modes at the cutoff scale $k = k_{\sigma}(N)$, which is already in the superhorizon regime. Altogether, this means that the only relevant non-Markovian effects will be generated from horizon crossing to coarse-grained crossing, a regime in which gradient terms are exponentially suppressed. Consequently, \Eq{phi UV compact} can be written as\footnote{This statement will be explicitly verified in section \ref{Sec:4}}

\begin{equation}
\dif{}{N} \left(a^3 H_\uir \epsilon_{1,\uir}\dif{}{N}\left(\frac{\phi_\uuv}{\pi_\uir}\right)\right) \simeq \mathcal{O}(\Xi_i),
\label{phi UV compact super}
\end{equation}
which has an analytical solution even in the presence of a IR background i.e.

\begin{align} 
    \phi_\uuv(N,k)&= C_a(k)\pi_\uir(N) + C_b(k)\pi_\uir(N)\int_0^{N} \frac{\dd N'}{a(N)^3 H_\uir(\phi_\uir(N),\pi_\uir(N)) \epsilon_{1,\uir}(\pi_\uir(N))} \label{phi UV solution}.
\end{align}

Initial conditions for the mode function \eqref{phi UV solution} are given by the Bunch–Davies vacuum, following the same prescription as in standard linear perturbation theory. Once \Eq{phi UV solution} is specified, it can be used to compute the variance of the noises by simply substituting it into \Eq{variance of noise}. Thus, under the assumptions made in this section, the stochastic system is fully characterized. From \Eq{phi UV solution}, one can clearly identify a Markovian term (proportional to $C_a(k)$), which depends only on the value of the IR variables at the instant $N$ at which the stochastic system is evaluated, and a non-Markovian term (proportional to $C_b(k)$), which depends on the entire history of the IR background from the time of horizon crossing of the first mode (which we set at $N=0$) up to $N$. 

We can now distinguish between attractor and non-attractor inflationary regimes by defining the second SR parameter as

\begin{equation}
    \epsilon_{2,\uir}\equiv \frac{1}{\epsilon_{1,\uir}}\dif{\epsilon_{1,\uir}}{N}.
\end{equation}

If $\epsilon_{2,\uir} > -3$, the term proportional to $C_b(k)$ decays exponentially, making non-Markovian effects negligible. We will refer to this as the attractor case, which corresponds to any inflationary regime where the curvature perturbation remains roughly constant on superhorizon scales (for example, slow-roll (SR)). Conversely, if $\epsilon_{2,\uir} < -3$, the non-Markovian term proportional to $C_b(k)$ becomes dominant. This implies that for non-attractor cases, in which the curvature perturbation grows on superhorizon scales, the stochastic system could be non-Markovian. Note that, even in non-attractor models, non-Markovianity does not necessarily arise if the integral appearing in the second term of \Eq{phi UV solution} can be reduced to a history-independent form. Notably, non-attractor regimes have recently attracted significant interest due to their phenomenological consequences, such as the generation of primordial black holes (PBHs) and scalar-induced gravitational waves (see \cite{Domenech:2021ztg, Escriva:2022duf} and references therein).

However, all the conclusions of this section rely heavily on the two assumptions made above, namely neglecting both $\mathcal{O}(k^2)$ corrections and terms in which the noises appear explicitly in the equation of motion for UV modes. In what follows, we will examine the validity of these assumptions in a perturbative regime, showing that while the former is perfectly justified, the same cannot be said for the latter.

\section{The perturbative stochastic inflationary system}\label{Sec:3}
In this section,  we analyze the stochastic formalism in a perturbative regime in order to obtain analytical solutions for the UV modes that solve \Eqs{phi UV taylor 2}-\eqref{pi UV taylor 2}, including all possible stochastic effects within the current formulation, \ie~not neglecting terms in which the noises appear explicitly in the equation of motion.

\subsection{Expansion of equations}
Following a similar technique as the one introduced in \cite{Cruces:2024pni}, we will assume that the amplitudes of the noise terms are much smaller than the IR quantities, as is indeed the case during inflation. In addition, we assume that these small quantities are of the same order as the perturbative parameters associated with $\phi_\uir$ and $\pi_\uir$, respectively. We therefore expand all equations of the system as
\begin{align}
    \phi_{\mathrm{IR}}(N) = \bar\phi + \phi_{\mathrm{IR}}^{(1)} + \phi_{\mathrm{IR}}^{(2)} + \cdots, \quad \pi_{\mathrm{IR}}(N) = \bar\pi + \pi_{\mathrm{IR}}^{(1)} + \pi_{\mathrm{IR}}^{(2)} + \cdots,\label{SNE phi IR}
\end{align}
\begin{align}
    \phi_\mathrm{UV}(N,k) = \phi_\mathrm{UV}^{(1)} + \phi_\mathrm{UV}^{(2)} + \cdots, \quad \pi_\mathrm{UV}(N,k) = \pi_\mathrm{UV}^{(1)} + \pi_\mathrm{UV}^{(2)}  + \cdots, \label{SNE phi UV}
\end{align}
Accordingly, the resulting perturbative equations are given below, together with the corresponding expansion of the noise terms:
\begin{align}
    \mathsf{N}_\phi
    &\equiv  \mathsf{N}^{(1)}_\phi +  \mathsf{N}^{(2)}_\phi  + \cdots,\quad \mathsf{N}_\pi\equiv  \mathsf{N}^{(1)}_\pi + \mathsf{N}^{(2)}_\pi\  + \cdots. \label{noise expansion}
\end{align}
In fact, one may also explicitly introduce the expansion parameters as follows. Equation~\eqref{variance of noise} gives rise to the dimensionless parameters
\begin{align}
    \lambda_\phi \equiv \cfrac{\sqrt{\left.\prn{\frac{k^3}{2\pi^2}\abs{\delta \phi_\mathrm{UV}(N,k;\bar\phi)}^2}\right|_{k=\bar k_\sigma(N)}}}{\bar\phi},\quad \lambda_\pi \equiv \cfrac{\sqrt{\left.\prn{\frac{k^3}{2\pi^2}\abs{\delta \pi_\mathrm{UV}(N,k;\bar\phi)}^2}\right|_{k=\bar k_\sigma(N)}}}{\bar\pi},
\end{align}
where $\bar k_\sigma(N)\equiv \sigma a \bar H$, and $\delta\phi_\mathrm{UV}$ and $\delta\pi_\uuv$ are solutions of the deterministic equations obtained by replacing the IR background $\phi_\uir, \pi_\uuv$ in the UV-mode equations \eqref{phi UV} and \eqref{pi UV} with the deterministic homogeneous background $\bar\phi,\bar \pi$, namely the usual Mukhanov--Sasaki equations in linear perturbation theory. Here, the background Hubble parameter $\bar H$ satisfies
\begin{align}
    \bar H = \sqrt{\frac{V(\bar\phi)}{3\mpl^2 - \frac{\bar \pi^2}{2}}}.
\end{align}
We can use the calculable deterministic initial values, $\lambda_{\phi,\mathrm{i}}\equiv \lambda_\phi(N=0)$ and $\lambda_{\pi,\mathrm{i}}\equiv \lambda_\pi(N=0)$ as expansion parameters. 

In the perturbative regime, the system at each order can be written as follows.
\begin{description}
    \item[0th order (homogeneous background)]\mbox{} \\
    At this order, the equations reduce simply to the homogeneous background equations:
\begin{align}
    \dif{\bar\phi}{N} &= \bar{\pi}, \label{phi background}\\
    \dif{\bar\pi}{N} &= -\left(3-\frac{\bar{\pi}^2}{2\mpl^2}\right)\left(\bar\pi+\mpl^2\frac{V_\phi(\bar\phi)}{V(\bar\phi)}\right) .
    \label{pi background}
\end{align} \label{0theq SNE}
    \item[1st order]\mbox{} \\
The first-order equations for the IR modes are given by
\begin{align}
    \dif{\phi_\uir^{(1)}}{N} &= {\pi_\uir^{(1)}} + \mathsf{N}_\phi^{(1)}\xi_\phi,  \label{phi IR first}\\
    \dif{\pi_\uir^{(1)}}{N} &= -\left(3-\epsilon_1 + \frac{\epsilon_1\epsilon_2}{3-\epsilon_1}\right)\pi_\uir^{(1)} \notag\\
    &\hspace{5mm}- \prn{-\frac{3}{2}\epsilon_2 + \frac{1}{2}\epsilon_1\epsilon_2 - \frac{1}{4}\epsilon_2^2 - \frac{1}{2}\epsilon_2\epsilon_3 + \frac{\epsilon_1\epsilon_2^2}{2(3-\epsilon_1)}}\phi_\uir^{(1)} + \mathsf{N}_\pi^{(1)}\xi_\pi,\label{pi IR first} 
\end{align}
where $\epsilon_1 \equiv -\dif{\ln \bar H}{N}$ and $\epsilon_n \equiv \dif{\epsilon_{n-1}}{N}$. The first-order UV modes obey the Mukhanov--Sasaki equations \footnote{Note that, unlike in Section \ref{Sec:2} ,where we used the IR background to rewrite the MS equation in a more compact way, we are now in a perturbative framework, meaning that we can use the fiducial background \eqref{phi background}–\eqref{pi background} to rewrite the MS equation.} \footnote{The difference between \Eq{pi IR first} and \Eq{pi UV first} in the $k\rightarrow 0$ limit can be typically neglected within the separate-universe assumption \cite{Cruces:2022imf}.},
\begin{align}
    0&=\pdif{\phi_\uuv^{(1)}}{N} - \pi_\uuv^{(1)} ,  \label{phi UV first}\\
    0&= \pdif{\pi_\uuv^{(1)}}{N} + \left(3-\epsilon_{1}\right)\pi_\uuv^{(1)} + \com{\frac{V_{,\phi\phi}(\bar\phi)}{\bar H^2} - \frac{1}{a^3\mpl^2 \bar H}\dif{}{N}\prn{a^3 \bar H \bar \pi^2}} \phi_\uuv^{(1)} + \frac{k^2}{a^2\bar H^2} \phi_\uuv^{(1)}\notag\\
    &= \pdif{\pi_\uuv^{(1)}}{N} + \left(3-\epsilon_{1}\right)\pi_\uuv^{(1)} + \prn{ -\frac{3}{2}\epsilon_{2} + \frac{1}{2}\epsilon_{1}\epsilon_{2} - \frac{1}{4}\epsilon_{2}^2 - \frac{1}{2}\epsilon_{2}\epsilon_{3} } \phi_\uuv^{(1)}+ \frac{k^2}{a^2\bar H^2}\phi_\uuv^{(1)},\label{pi UV first}
\end{align}
and, together with the Bunch--Davies vacuum condition in the infinite past, allow one to compute the amplitudes of the noise terms from deterministic quantities:
\begin{align}
    \mathsf{N}^{(1)}_\phi \equiv \sqrt{\left.\prn{\prn{1-\epsilon_{1}}\frac{k^3}{2\pi^2}\abs{\phi_\uuv^{(1)}(N,k)}^2}\right|_{k=\bar k_\sigma(N)}},\quad \mathsf{N}^{(1)}_\pi \equiv \sqrt{\left.\prn{\prn{1-\epsilon_{1}}\frac{k^3}{2\pi^2}\abs{\pi_\uuv^{(1)}(N,k)}^2}\right|_{k=\bar k_\sigma(N)}}.
\end{align}
For later convenience, let us rewrite these equations in matrix form:
\begin{align}
    &\dif{}{N}\mathbf{\Phi}^{(1)}_\mathrm{IR} + \widetilde{\mathbf{A}}\mathbf{\Phi}^{(1)}_\mathrm{IR} =\mathbf{\Xi}^{(1)}, \label{IR modes 1st}\\
    &\pdif{}{N}\mathbf{\Phi}^{(1)}_\mathrm{UV} + \mathbf{A}\mathbf{\Phi}^{(1)}_\mathrm{UV} =0 \label{UV modes 1st}
\end{align} 
where 
\begin{align}
   \mathbf{\Phi}^{(n)}_\mathrm{IR}(N,k;\bar\phi,\bar\pi) \equiv \left[\begin{array}{c}
         \phi_\mathrm{IR}^{(n)}  \\
       \pi_\mathrm{IR}^{(n)}   
    \end{array}\right],\quad   \mathbf{\Phi}^{(n)}_\mathrm{UV}(N,k;\bar\phi,\bar\pi) \equiv \left[\begin{array}{c}
         \phi_\mathrm{UV}^{(n)}  \\
       \pi_\mathrm{UV}^{(n)}   
    \end{array}\right],
\end{align}
and 
\begin{align}
    \mathbf{\Xi}^{(n)}(N) \equiv \left[\begin{array}{c}
         \mathsf{N}_\phi^{(n)}\xi_\phi  \\
       \mathsf{N}_\pi^{(n)}\xi_\pi   
    \end{array}\right].
\end{align}
Note that the following relation holds within the validity of the separate-universe assumption adopted throughout this paper:
\begin{align}
    \mathbf{A}(N,k,\bar\phi,\bar\pi) \simeq \widetilde{\mathbf{A}} (N,\bar\phi,\bar\pi) + \mathbf{W}(N,k) \label{def A},
\end{align}
where
\begin{align}
    {\mathbf{W}}(N,k;\bar\phi,\bar\pi) \equiv \left[\begin{array}{cc}
         0 & 0  \\
        \frac{k^2}{a^2\bar H^2} & 0
    \end{array}\right]. \label{def W UV}
\end{align}

    \item[2nd order]\mbox{} \\
    In matrix form, the equations for the IR and UV modes are given by
\begin{align}
    &\dif{}{N}\mathbf{\Phi}^{(2)}_\mathrm{IR} + \widetilde{\mathbf{A}}\mathbf{\Phi}^{(2)}_\mathrm{IR} = {\mathbf{B}_\uir^{(1)}}\mathbf{\Phi}^{(1)}_\mathrm{IR} + \mathbf{\Xi}^{(2)},\label{IR modes 2nd}\\
    &\pdif{}{N}\mathbf{\Phi}^{(2)}_\mathrm{UV} + \mathbf{A}\mathbf{\Phi}^{(2)}_\mathrm{UV} =\mathbf{B}_\uuv^{(1)}\mathbf{\Phi}^{(1)}_\mathrm{IR}, \label{UV modes 2nd}
\end{align} 
where 
\begin{align}
    \mathbf B_\uir^{(1)} &\equiv  -\frac{\partial \widetilde{\mathbf{A}}}{\partial \bar{\phi}}\phi_\mathrm{IR}^{(1)}-  \frac{\partial \widetilde{\mathbf{A}}}{\partial \bar{\pi}}\pi_\mathrm{IR}^{(1)}, \label{def B IR}\\
 \mathbf B_\uuv^{(1)} &\equiv  -\frac{\partial {\mathbf{A}}}{\partial \bar{\phi}}\phi_\mathrm{UV}^{(1)}-  \frac{\partial {\mathbf{A}}}{\partial \bar{\pi}}\pi_\mathrm{UV}^{(1)}= -\frac{\partial \widetilde{\mathbf{A}}}{\partial \bar{\phi}}\phi_\mathrm{UV}^{(1)}-  \frac{\partial \widetilde{\mathbf{A}}}{\partial \bar{\pi}}\pi_\mathrm{UV}^{(1)} -\frac{\partial {\mathbf{W}}}{\partial \bar{\phi}}\phi_\mathrm{UV}^{(1)}-  \frac{\partial {\mathbf{W}}}{\partial \bar{\pi}}\pi_\mathrm{UV}^{(1)}.\label{def B UV}
\end{align}
The second-order UV solutions, together with the first-order ones, give rise to the noise amplitudes $\mathsf{N}_\phi^{(2)}$ and $\mathsf{N}_\pi^{(2)}$, which can be decomposed into two parts:
\begin{align}
    \mathsf{N}^{(2)}_\phi ={}_\mathrm{det}\mathsf{N}^{(2)}_\phi + {}_{\mathrm{sto}}\mathsf{N}^{(2)}_\phi,\quad \mathsf{N}^{(2)}_\pi ={}_\mathrm{det}\mathsf{N}^{(2)}_\pi + {}_{\mathrm{sto}}\mathsf{N}^{(2)}_\pi.
\end{align}

The first part (labeled ``det") corresponds to the standard functional Taylor expansion of the result obtained in Section \ref{Sec:2} (see \App{Sec:A} for an explicit calculation):

\begin{align}
    {}_\mathrm{det}\mathsf{N}^{(2)}_\phi &\equiv \int_0^N\dd N'\left[\frac{\delta \mathsf{N}^{(1)}_\phi}{\delta \bar{\phi}(N')} \phi_{\mathrm{IR}}^{(1)}(N')+\frac{\delta \mathsf{N}^{(1)}_\phi}{\delta \bar{\pi}(N')} \pi_{\mathrm{IR}}^{(1)}(N')\right], \label{def Nphi det}\\
    {}_\mathrm{det}\mathsf{N}^{(2)}_\pi &\equiv \int_0^N\dd N'\left[\frac{\delta \mathsf{N}^{(1)}_\pi}{\delta \bar{\phi}(N')} \phi_{\mathrm{IR}}^{(1)}(N')+\frac{\delta \mathsf{N}^{(1)}_\pi}{\delta \bar{\pi}(N')} \pi_{\mathrm{IR}}^{(1)}(N')\right], \label{def Npi det}
\end{align}
Note that the functional derivative acts both on $\phi_\uuv^{(1)}(N,k;\bar\phi,\bar\pi)$ itself and on the background fields appearing in $k_\sigma(N)$ after the evaluation at $k=k_\sigma(N)$. The second part (labeled ``sto"), by contrast, corresponds to a term whose functional form cannot be computed deterministically, but instead arises from stochastic noise contributions that were neglected in Section \ref{Sec:2}. It can be expressed as
\begin{align}
    &{}_{\mathrm{sto}}\mathsf{N}^{(2)}_\phi= \sqrt{\left.\prn{\prn{1-\epsilon_{1}}\frac{k^3}{2\pi^2}\abs{{}_\mathrm{sto}\phi_\uuv^{(2)}(N,k)}^2}\right|_{k=\bar k_\sigma(N)}}, \label{def Nphi sto}\\
    &{}_{\mathrm{sto}}\mathsf{N}^{(2)}_\pi= \sqrt{\left.\prn{\prn{1-\epsilon_{1}}\frac{k^3}{2\pi^2}\abs{{}_\mathrm{sto}\pi_\uuv^{(2)}(N,k)}^2}\right|_{k=\bar k_\sigma(N)}}. \label{def Npi sto}
\end{align}

Similarly, the second-order UV solutions are decomposed as
\begin{align}
    \phi_\uuv^{(2)}\equiv \udet\phi_\uuv^{(2)} + \usto\phi_\uuv^{(2)},\quad \pi_\uuv^{(2)}\equiv \udet\pi_\uuv^{(2)} + \usto\pi_\uuv^{(2)}.
\end{align}
Here, the deterministic part of the UV modes, whose contribution is included in $\udet\mathsf{N}_\phi^{(2)}$ and $\udet\mathsf{N}_\pi^{(2)}$, is simply the first order term of the functional Taylor expansion of \Eq{phi UV solution}, defined by
\begin{align}
\,_\mathrm{det}\mathbf{\Phi}_{\mathrm{UV}}^{(2)}\equiv \int_0^N\dd N'\left[\frac{\delta \mathbf{\Phi}_{\mathrm{UV}}^{(1)}[\bar\phi,\bar\pi;N]}{\delta \bar{\phi}(N')} \phi_{\mathrm{IR}}^{(1)}(N')+\frac{\delta \mathbf{\Phi}_{\mathrm{UV}}^{(1)}[\bar\phi,\bar\pi;N]}{\delta \bar{\pi}(N')} \pi_{\mathrm{IR}}^{(1)}(N')\right], \label{def UV second det}
\end{align}
where $\udet\mathbf{\Phi}_\uuv^{(2)}=(\udet\phi_\uuv^{(2)},\udet\pi_\uuv^{(2)})$. This follows from the fact that standard perturbation theory for a functional implies that the second-order perturbation $\mathbf{\Phi}_{\mathrm{UV}}^{(2)}$ is given by the convolution of the functional derivatives of the first-order solution $\mathbf{\Phi}_{\mathrm{UV}}^{(1)}$ with respect to $\bar{\phi}$ and $\bar{\pi}$ over the history of the IR perturbations $\phi_{\mathrm{IR}}^{(1)}$ and $\pi_{\mathrm{IR}}^{(1)}$. We then have that

\begin{equation}
    \sum_{i=1}^{\infty}\,_\mathrm{det}\mathbf{\Phi}_{\mathrm{UV}}^{(i)}=\,_\mathrm{det}\mathbf{\Phi}_{\mathrm{UV}},
\end{equation}
where $\,_\mathrm{det}\mathbf{\Phi}_{\mathrm{UV}} = (\udet\phi_\uuv, \udet\pi_\uuv)$, with $\udet\phi_\uuv$ given by \Eq{phi UV solution} and $\udet\pi_\uuv$ its corresponding time derivative.

\end{description}
One can derive the equations at higher orders in the same manner. By construction, summing the equations at all orders reproduces the original equations. For example,
\begin{align}
    \text{Eq.~\eqref{phi UV first}} + \text{Eq.~\eqref{UV modes 2nd}} + \cdots = \text{Eq.~\eqref{phi UV taylor 2}},
\end{align}
where no $\mathcal{O}(\Xi_i)$ terms have been neglected.

Let us also note that the perturbative quantities satisfy the following initial conditions:
\begin{align}
    &\mathbf\Phi_\uir^{(n)}(0) = 0\quad \mathrm{for}\ n\geq1, \label{initial conditions IR}\\
    &\mathbf\Phi_\uuv^{(1)}(0,k) = \mathbf\Phi_\mathrm{ini}(k),\quad \mathbf\Phi_\uuv^{(n)}(0,k) = 0\quad \mathrm{for}\ n\geq2,\label{initial conditions UV}
\end{align}
where the initial time is taken to be $N=0$, corresponding to the onset of inflation, when the coarse-grained Hubble patch $k_\sigma^{-1}$ corresponds to the entire observable universe. These conditions follow from the fact that no stochastic process acts on our observable scales before this onset. In particular, the first-order UV-mode equation is solved with the initial condition $\mathbf\Phi_\mathrm{ini}(k)$ associated with the Bunch--Davies vacuum, whereas the higher-order nonlinear UV-mode equations are solved with vanishing initial values.

\subsection{Perturbative solutions for the IR and UV modes}
First, let us assume that the homogeneous background equations can be solved, so that the background quantities are known as functions of time, \ie\ $\bar\phi=\bar\phi(N)$ and $\bar\pi=\bar\pi(N)$. Under this assumption, the UV-mode equations at each order can be solved formally owing to their linearity. With the initial conditions \eqref{initial conditions UV}, the solutions to \eqref{UV modes 1st} and \eqref{UV modes 2nd} can be written as
\begin{align}
    \mathbf{\Phi}_\mathrm{UV}^{(1)}(N,k) &= \mathbf{U}(N,0)\mathbf{\Phi}_\mathrm{ini}(k),
    \label{solution_UV_first}\\
    \mathbf{\Phi}_\mathrm{UV}^{(2)}(N,k) &= \int_0^N \dd N' \mathbf{U}(N,N')\mathbf{B}_\uuv^{(1)}(N',k)\mathbf{\Phi}^{(1)}_\mathrm{IR}(N'),
    \label{solution_UV_second}
\end{align}
where the time-ordered exponential $\mathbf{U}$ is defined by
\begin{align}
    \mathbf{U}(N,N')\equiv\mathrm{T}\exp \left(- \int_{N'}^N \dd u {\mathbf{A}}(u)\right), \label{def U}
\end{align}
and $\mathbf{U}^{-1}$ denotes its inverse.

In addition to these formal solutions, we will also make use of the analytical large-scale solution for $\phi_{\uuv}^{(1)}$, which, within the separate universe approach, takes the same form as \Eq{phi UV solution} by substituting the IR background by the fiducial global background of \Eqs{phi background}-\eqref{pi background}
\begin{align} 
    \phi_{\mathrm{UV}}^{(1)}(N,k)&= C_a^{(1)}(k)\bar{\pi}(N) + C_b^{(1)}(k)\bar{\pi}(N)\int_0^{N} \dd N' f(N')\label{solution_SUA},
\end{align}
where
\begin{equation}
    f(N)\equiv 2\mpl^2\frac{1}{a(N)^3\bar{\pi}(N)^2 H(\bar{\phi}(N),\bar{\pi}(N))}.
    \label{def_F}
\end{equation}
Note that, in the case of a scalar field evolving on a fixed de Sitter spacetime with $H=\mathrm{const.}$, $f$ is given by $f_{\mathrm{dS}}(N)\equiv 2\mpl^2/\prn{a^3\bar{\pi}^2 H}$. For the purpose of computing the noise terms, the coefficients of the independent solutions, $C^{(1)}_i(k)$, are evaluated at the coarse-graining scale and may therefore be regarded as $C^{(1)}_i(k=\bar k_\sigma(N))= C^{(1)}_i(\sigma a(N)H[\bar\phi,\bar\pi])$. This provides a definition of $\phi_\mathrm{UV}^{(1)}$ as a functional of $\bar\phi$ and $\bar\pi$, with explicit time dependence through $a(N)$, \ie\ $\phi_\mathrm{UV}^{(1)}[\bar\phi,\bar\pi;N]$.

\section{Analysis for non-Markovianity}\label{Sec:4}

Up to first order in perturbation theory, the IR system is Markovian, since the solution \eqref{UV modes 1st}, or equivalently \eqref{solution_SUA}, is deterministic, and therefore the amplitudes of the first-order noise terms are independent of stochasticity. Then, non-Markovianity might appear only at second and higher orders, as already implied in the previous section. 

In what follows, we derive and discuss analytical expressions for the non-Markovian terms. To clarify how the noise terms are affected by the IR modes through the UV-mode solutions, let us rewrite the second-order solution \eqref{solution_UV_second} in terms of its first-order counterparts for both the IR and UV modes, and classify the resulting contributions in the manner introduced in the previous section: the part whose functional form can be determined through the functional derivative \eqref{def UV second det}, and the part that cannot. Under this classification, the analysis of non-Markovian contributions becomes straightforward, particularly when examining their scale dependence or identifying a non-Markovian term as an integral over the IR quantities.

Let us consider a general attractor case in order to gain clear insight into the structure without cumbersome calculations (the non-attractor case is studied in \App{Sec:B.2}). In this case, only the first term is relevant, as in slow-roll inflation, and hence
\begin{equation}
    \mathbf{\Phi}^{(1)}_\mathrm{UV}(N,k)\simeq C_a^{(1)}(k)\left[\begin{array}{c}
         \bar{\pi}(N)  \\
       \frac{d\bar{\pi}(N)}{dN}  
    \end{array}\right]
    \label{solution_SUA_att}.
\end{equation}
Through the algebraic manipulations presented in \App{Sec:B.1}, the second-order UV-mode solution can be expressed as
\begin{align} 
   \mathbf{\Phi}_{\mathrm{UV}}^{(2)}(N,k)    =& \,_\mathrm{det}\mathbf{\Phi}_{\mathrm{UV}}^{(2)}(N,k)  + \,_\mathrm{sto}\mathbf{\Phi}_{\mathrm{UV}}^{(2)}(N,k),
   \label{solution_UV_second_final}
\end{align}
where
\begin{align}
   \,_\mathrm{det}\mathbf{\Phi}_{\mathrm{UV}}^{(2)}(N,k) &= -C_a^{(1)}(k){\mathbf{A}}(N,k)\mathbf{\Phi}_\mathrm{IR}^{(1)}(N) ,\label{phi UV second det ori}\\
   \,_\mathrm{sto}\mathbf{\Phi}_{\mathrm{UV}}^{(2)}(N,k) &=C_a^{(1)}(k)\int_0^{N}\dd N'\mathbf{U}(N,N'){\mathbf{A}}(N',k)\mathbf{\Xi}^{(1)}(N')\notag\\
   &\quad+ C_a^{(1)}(k)  \int_0^{N}\dd N'\mathbf{U}(N,N'){\mathbf{A}}(N',k){\mathbf{W}}(N',k)\mathbf{\Phi}_\mathrm{IR}^{(1)}(N').\label{phi UV second sto ori}
\end{align}
When evaluating the noise terms using these solutions, the UV modes are to be evaluated at the coarse-graining scale, \ie\ at $k=k_\sigma(N)$. Consequently, at leading order in the gradient expansion, the solutions can be rewritten as
\begin{align}
    \,_\mathrm{det}\mathbf{\Phi}_{\mathrm{UV}}^{(2)}(N,k_\sigma(N)) &= -C_a^{(1)}(k)\widetilde{\mathbf{A}}(N)\mathbf{\Phi}_\mathrm{IR}^{(1)}(N) ,\label{phi UV second det}\\
   \,_\mathrm{sto}\mathbf{\Phi}_{\mathrm{UV}}^{(2)}(N,k_\sigma(N)) &=C_a^{(1)}(k)\int_0^{N}\dd N'\mathbf{U}(N,N'){\mathbf{A}}(N',k_\sigma(N))\mathbf{\Xi}^{(1)}(N')\notag\\
   &\quad+ C_a^{(1)}(k)  \int_0^{N}\dd N'\mathbf{U}(N,N'){\mathbf{A}}(N',k_\sigma(N)){\mathbf{W}}(N',k_\sigma(N))\mathbf{\Phi}_\mathrm{IR}^{(1)}(N').\label{phi UV second sto}
\end{align}
These expressions make the structure of the noise terms transparent. In particular, they show that the contributions can be separated into two parts, each of which admits a distinct interpretation, as explained below.

The first term corresponds to the part studied in Section \ref{Sec:2}. In the language of this section, it  can be determined once the functional dependence of the deterministic first-order UV solution on $\bar\phi$ and $\bar\pi$ has been specified, so that functional derivatives of arbitrary order can be taken. This is why the notation ``det'' is used for the first part. Indeed, this term can be shown to be equivalent to its definition in \Eq{def UV second det}, by using the attractor solution \eqref{solution_SUA_att} and its functional derivatives:
\begin{align} 
    \frac{\delta \mathbf{\Phi}_{\mathrm{UV}}^{(1)}(N,k_\sigma(N))}{\delta \bar{\phi}(N')}&=C_a^{(1)}(k_\sigma(N))\delta(N-N')\left[\begin{array}{c}
         0  \\
       \frac{\partial}{\partial \bar{\phi}}\left(\frac{d\bar{\pi}}{dN'}\right)  
    \end{array}\right]=-C_a^{(1)}(k_\sigma(N))\delta(N-N')\left[\begin{array}{c}
         0  \\
       \widetilde{A}_{21} 
    \end{array}\right],\label{functional derivatieves phi}\\
    \frac{\delta \mathbf{\Phi}_{\mathrm{UV}}^{(1)}(N,k_\sigma(N))}{\delta \bar{\pi}(N')}&=C_a^{(1)}(k_\sigma(N))\delta(N-N')\left[\begin{array}{c}
         1  \\
       \frac{\partial}{\partial \bar{\pi}}\left(\frac{d\bar{\pi}}{dN'}\right)  
    \end{array}\right]=-C_a^{(1)}(k_\sigma(N))\delta(N-N')\left[\begin{array}{c}
         -1  \\
       \widetilde{A}_{22}  
    \end{array}\right].
    \label{functional_derivatives_att}
\end{align}
As anticipated in Section \ref{Sec:2}, the important point is that this term involves no memory effect. Therefore, in the attractor case, it corresponds to a Markovian contribution, for which only the value of the IR modes at the instantaneous time $N$ is required.

On the other hand, the second part, denoted by ``sto'' in \Eq{solution_UV_second_final}, explicitly represents history dependence, \ie\ non-Markovianity, through integrals over the IR quantities. Let us examine the contributions from these terms for a fixed time $N$ and the corresponding mode at the cutoff scale, $k_\sigma(N)$. We will first focus on the second term of \Eq{phi UV second sto}. First, in the regime where the UV modes are deep inside the horizon, the $k$-dependent term $\mathbf{W}$ dominates the propagator $\mathbf{U}(N,N')$, leading to an exponential suppression of these modes according to the definition of the propagator in \Eq{def U}. This is consistent with the expectation discussed in \Sec{Sec:2}: the characteristic time scale of vacuum oscillations for modes deep inside the horizon is much shorter than that of the IR modes, and therefore non-Markovianity does not play an important role. Second, let us consider the opposite regime, in which the mode is close to the coarse-graining scale. In this case, since the noise terms are evaluated at the coarse-graining scale $k_\sigma(N)=\sigma a H$, this contribution is much smaller than the Markovian term \eqref{phi UV second det}, because it is suppressed by $\mathbf{W}$, which is itself of order $\sigma$. Therefore, at leading order in the gradient expansion, the second line of \Eq{phi UV second sto} can be safely neglected, justifying the assumption in Section \ref{Sec:2} of dropping the $\mathcal{O}(k^2)$ terms in the equations of motion for UV modes.

We will now turn our attention to the first line of \Eq{phi UV second sto}, which originates from $\mathcal{O}(\Xi_i)$ terms in \Eq{phi UV compact}. In the standard construction of stochastic inflation, it is assumed to be negligible when evaluating the noise in the second-order Langevin equation for the IR modes. The reason is that this contribution takes the form of a term quadratic in the noise, such quantities are discarded from the beginning. The main objective of this manuscript is to check the validity of this assumption. To do so, we will study the simplified case in which $\tilde{\mathbf{A}}$ is a constant. Neglecting $\mathcal{O}(\sigma^2)$ terms in \Eq{phi UV second sto} we have

\begin{align} \notag
\,_\mathrm{sto}\mathbf{\Phi}_{\mathrm{UV}}^{(2)}(N,k) &\simeq C_a^{(1)}(k)\int_0^{N}\dd N'\mathbf{U}(N,N')\tilde{\mathbf{A}}\mathbf{\Xi}^{(1)}(N')\\ \notag
&\simeq C_a^{(1)}(k)\int_0^{N}\dd N'\mathrm{T}\exp \left(- \tilde{\mathbf{A}}\int_{N'}^N \dd u \right)\tilde{\mathbf{A}}\mathbf{\Xi}^{(1)}(N')\\ 
&\simeq C_a^{(1)}(k)\tilde{\mathbf{A}}\mathbf{\Phi}_\mathrm{IR}^{(1)}(N)\,,
\label{phi UV second sto Aconstant}
\end{align}
where we have used that $\mathbf U(N,N')$ commutes with a constant $\tilde{\mathbf{A}}$ and the solution of the leading order stochastic equation \eqref{IR modes 1st} for $\mathbf{\Phi}_\mathrm{IR}^{(1)}(N)$. 

As we can clearly see, the deterministic part of the second-order UV-mode solution $\,_\mathrm{det}\mathbf{\Phi}_{\mathrm{UV}}^{(2)}$ of \Eq{phi UV second det ori} for constant $\tilde{\mathbf{A}}$ is exactly the same as its stochastic counterpart $\,_\mathrm{sto}\mathbf{\Phi}_{\mathrm{UV}}^{(2)}$ of \Eq{phi UV second sto Aconstant} but with an overall minus sign. The total second-order UV-mode solution for constant $\tilde{\mathbf{A}}$ is

\begin{equation}
    \mathbf{\Phi}_{\mathrm{UV}}^{(2)}(N,k)    =\,_\mathrm{det}\mathbf{\Phi}_{\mathrm{UV}}^{(2)}(N,k)  + \,_\mathrm{sto}\mathbf{\Phi}_{\mathrm{UV}}^{(2)}(N,k) \simeq 0,
    \label{phi UV second full Aconstant}
\end{equation}

Note that \Eq{phi UV second full Aconstant} makes perfect sense. Indeed, a constant $\tilde{\mathbf{A}}$ corresponds to a free field in de-Sitter, which is perfectly Gaussian. Since non-Gaussianities are proportional to $\mathbf{\Phi}_{\mathrm{UV}}^{(2)}(N,k)$, we expect \Eq{phi UV second full Aconstant} to hold in this case.  The key lesson is that terms which naively contribute as second-order noises in the second-order stochastic equation of motion for $\mathbf{\Phi}_{\mathrm{IR}}^{(2)}(N)$ are not only of the same perturbative order as terms that apparently appear linear in the noises, but are also essential for obtaining reliable physical results. A similar conclusion has been reached recently in \cite{Palma:2023idj,Palma:2023uwo,Palma:2025oux}. In a companion paper, we will explore the issue of nonlinear noises in greater detail.

\section{Discussion and summary}\label{Sec:5}

The stochastic formalism is characterized by both the Langevin equations for the IR modes, \eqref{phi IR} and \eqref{pi IR}, and the equations for the UV modes, \eqref{phi UV} and \eqref{pi UV}, which determine the noise terms appearing in the Langevin equations. As discussed in \Sec{Sec:2}, the non-Markovianity intrinsic to this system generally arises because the same field is separated into different scales. Most previous works have assumed Markovianity without a quantitative assessment of its validity. It is therefore important to provide an analytical justification for this assumption. In this work, we have analytically derived the non-Markovian terms of the stochastic system in a perturbative regime.

We formulated a perturbative expansion in order to clarify the effects of non-Markovianity. In particular, the first order contains only Markovian contributions, since the noise terms are evaluated on top of the homogeneous background. By contrast, the second order contains two parts, which we refer to as the ``deterministic'' and ``stochastic'' contributions. The former is associated with the Taylor-like functional expansion, in which the functional form of the noise terms is determined by replacing the homogeneous background with the IR background in the first-order UV-mode solutions, \ie\ in the solutions to the Mukhanov--Sasaki equations. The latter can be understood as the deviation from the former arising from the presence of stochastic noise terms in the UV-mode equations.

We explicitly demonstrate the decomposition of the second-order UV-mode solution into the ``deterministic'' and ``stochastic'' parts defined above by directly solving the UV-mode equations at both first and second order and expressing the second-order solutions in terms of the first-order ones. These second-order solutions act as sources for the noise terms evaluated at the coarse-graining scale, as shown in \Eqs{phi UV second det}, \eqref{phi UV second sto}, \eqref{phi UV second det nonattractor}, and \eqref{phi UV second sto nonattractor}. Our main results are as follows. Under the standard assumption of neglecting terms quadratic in noises, we find that the UV-modes contribute to the IR system as a Markovian noise term in the attractor case, whereas it generically becomes a non-Markovian noise term in the presence of non-attractor phases. This contribution corresponds to the deterministic part of \Eqs{phi UV second det} and \eqref{phi UV second det nonattractor}, which is nothing more than a perturbative expansion of the naive analysis performed in Section \ref{Sec:2}. We then turn our attention to the stochastic part of the noises, which have been systematically neglected in the literature, and find that they are of the same order as the deterministic counterpart, in fact, we find that for constant $\tilde{\mathbf{A}}$, the stochastic and deterministic parts are given by exactly the same expression, but with opposite sign. This simple example demonstrates that terms quadratic in the noises, far from being negligible, are of precisely the same perturbative order as those typically studied, and are essential for the consistency of the formalism.

Our results establish that a fully consistent treatment of stochastic inflation beyond leading order must include the terms quadratic in the noises that appear in the equations of motion for the UV modes. In a companion paper, we will use the results of the present analysis to compute equal-time correlators of IR modes with all quadratic noise contributions properly taken into account.

\section*{Acknowledgments}
This work is supported by the National Key Research and Development Program of China
Grant No. 2021YFC2203004.  D.C. is supported by the
78th Batch of General Grants of the China Postdoctoral Science Foundation No. 2025M783447. T.K. acknowledges financial support from the Institute for Basic Science (IBS) under the project code, IBS-R018-D3 and JST SPRING, Japan Grant Number JPMJSP2180. The authors would also like to thank Cristiano Germani, Gonzalo Palma, Spyros Sypsas, Guillermo Ballesteros, Gerasimos Rigopoulos and Masahide Yamaguchi for useful discussions.

\appendix

\section{Deterministic part of the noise amplitudes}\label{Sec:A}
Let us explicitly demonstrate the calculation of the deterministic part of the second-order noise amplitude defined in \Eq{def Nphi det}. The corresponding discussion for the momentum term proceeds in exactly the same way.

In the expansion of $\mathsf{N}_\phi$ in \Eq{noise expansion}, the explicit expression is obtained by using the expansions $k_\sigma(N)=\bar k_\sigma(N) +  k^{(1)}_\sigma(N) + \cdots$ and $\epsilon_{1,\mathrm{IR}}= \epsilon_1 + \epsilon_1^{(1)} + \cdots$:
\begin{align}
    &\mathsf{N}^{(2)}_\phi(N) \notag\\
    &\equiv \sqrt{\prn{1 -\epsilon_{1}}\frac{\bar k_\sigma^3}{2\pi^2}\abs{ \phi_\mathrm{UV}^{(2)}(N,\bar k_\sigma(N))}^2} +  \sqrt{\prn{1 -\epsilon_{1}}\frac{\bar k_\sigma^3}{2\pi^2}}\pdif{\abs{ \phi_\mathrm{UV}^{(1)}(N,\bar k_\sigma(N))}}{k} k^{(1)}_\sigma \notag\\
    &\quad+ \prn{\frac{3}{2}\frac{k_\sigma^{(1)}}{\bar k_\sigma} + \frac{1}{2}\frac{\epsilon_1^{(1)}}{{1-\epsilon_{1}}}}\sqrt{\prn{1-\epsilon_{1}}\frac{\bar k_\sigma^3}{2\pi^2}\abs{ \phi_\mathrm{UV}^{(1)}(N,\bar k_\sigma(N))}^2} \notag\\
    &= \sqrt{\prn{1 -\epsilon_{1}}\frac{\bar k_\sigma^3}{2\pi^2}\abs{ \phi_\mathrm{UV}^{(2)}(N,\bar k_\sigma(N))}^2}   \notag\\
    &\quad + \sqrt{(1-\epsilon_1)\frac{\bar k_\sigma^3}{2\pi^2}}\int_0^N\dd N'\left[\left.\frac{\delta \phi_\mathrm{UV}^{(1)}(N,\bar k_\sigma(N))}{\delta \bar{\phi}(N')}\right|_{N\ \mathrm{fixed}} \phi_{\mathrm{IR}}^{(1)}(N')+\left.\frac{\delta \phi_\mathrm{UV}^{(1)}(N,\bar k_\sigma(N))}{\delta \bar{\pi}(N')}\right|_{N\ \mathrm{fixed}} \pi_{\mathrm{IR}}^{(1)}(N')\right]\notag\\
    &\quad+ \abs{ \phi_\mathrm{UV}^{(1)}(N,\bar k_\sigma(N))} \times \int_0^N\dd N'\left[\frac{\delta \sqrt{\prn{1-\epsilon_{1}}\frac{\bar k_\sigma^3}{2\pi^2}} }{\delta \bar{\phi}(N')} \phi_{\mathrm{IR}}^{(1)}(N')+\frac{\delta \sqrt{\prn{1-\epsilon_{1}}\frac{\bar k_\sigma^3}{2\pi^2}}}{\delta \bar{\pi}(N')} \pi_{\mathrm{IR}}^{(1)}(N')\right]\notag\\ 
    &= \sqrt{\prn{1 -\epsilon_{1}}\frac{\bar k_\sigma^3}{2\pi^2}\abs{ \phi_\mathrm{UV}^{(2)}(N,\bar k_\sigma(N))}^2} \notag\\
    &\quad+ \prn{\frac{\pdif{\abs{ \phi_\mathrm{UV}^{(1)}(N,\bar k_\sigma(N))}}{k}}{\abs{ \phi_\mathrm{UV}^{(1)}(N,\bar k_\sigma(N))}}k_\sigma^{(1)} +  \frac{3}{2}\frac{k_\sigma^{(1)}}{\bar k_\sigma} + \frac{1}{2}\frac{\epsilon_1^{(1)}}{{1-\epsilon_{1}}}}\sqrt{\prn{1 -\epsilon_{1}}\frac{\bar k_\sigma^3}{2\pi^2}\abs{ \phi_\mathrm{UV}^{(1)}(N,\bar k_\sigma(N))}^2}. 
\end{align}
Focusing on the deterministic part involving $\udet\phi_\uuv^{(2)}$, one can show that \Eq{def Nphi det} holds:
\begin{align}
    &\sqrt{(1-\epsilon_1)\frac{\bar k_\sigma^3}{2\pi^2}|{}_{\mathrm{det}}\phi_\mathrm{UV}^{(2)}(N,\bar k_\sigma(N))|^2}  +  \sqrt{\prn{1 -\epsilon_{1}}\frac{\bar k_\sigma^3}{2\pi^2}}\pdif{\abs{ \phi_\mathrm{UV}^{(1)}(N,\bar k_\sigma(N))}}{k} k^{(1)}_\sigma\notag\\
    &\quad + \prn{\frac{3}{2}\frac{k_\sigma^{(1)}}{\bar k_\sigma} + \frac{1}{2}\frac{\epsilon_1^{(1)}}{{1-\epsilon_{1}}}}\sqrt{(1 - \epsilon_1)\frac{k_\sigma^3}{2\pi^2}|\phi_\mathrm{UV}^{(1)}(k_\sigma)|^2}\notag\\
    &=\sqrt{(1-\epsilon_1)\frac{\bar k_\sigma^3}{2\pi^2}}\int_0^N\dd N'\left[\left.\frac{\delta \phi_\mathrm{UV}^{(1)}(N,\bar k_\sigma(N))}{\delta \bar{\phi}(N')}\right|_{k\ \mathrm{fixed}} \phi_{\mathrm{IR}}^{(1)}(N')+\left.\frac{\delta \phi_\mathrm{UV}^{(1)}(N,\bar k_\sigma(N))}{\delta \bar{\pi}(N')}\right|_{k\ \mathrm{fixed}} \pi_{\mathrm{IR}}^{(1)}(N')\right] \notag\\
    &\quad + \sqrt{(1-\epsilon_1)\frac{\bar k_\sigma^3}{2\pi^2}}\int_0^N\dd N'\left[\left.\frac{\delta \phi_\mathrm{UV}^{(1)}(N,\bar k_\sigma(N))}{\delta \bar{\phi}(N')}\right|_{N\ \mathrm{fixed}} \phi_{\mathrm{IR}}^{(1)}(N')+\left.\frac{\delta \phi_\mathrm{UV}^{(1)}(N,\bar k_\sigma(N))}{\delta \bar{\pi}(N')}\right|_{N\ \mathrm{fixed}} \pi_{\mathrm{IR}}^{(1)}(N')\right]\notag\\
    &\quad+ \abs{ \phi_\mathrm{UV}^{(1)}(N,\bar k_\sigma(N))} \times \int_0^N\dd N'\left[\frac{\delta \sqrt{\prn{1-\epsilon_{1}}\frac{\bar k_\sigma^3}{2\pi^2}} }{\delta \bar{\phi}(N')} \phi_{\mathrm{IR}}^{(1)}(N')+\frac{\delta \sqrt{\prn{1-\epsilon_{1}}\frac{\bar k_\sigma^3}{2\pi^2}}}{\delta \bar{\pi}(N')} \pi_{\mathrm{IR}}^{(1)}(N')\right]\notag\\ 
    &= \int_0^N\dd N'\left[\frac{\delta \mathsf{N}^{(1)}_\phi}{\delta \bar{\phi}(N')} \phi_{\mathrm{IR}}^{(1)}(N')+\frac{\delta \mathsf{N}^{(1)}_\phi}{\delta \bar{\pi}(N')} \pi_{\mathrm{IR}}^{(1)}(N')\right] = {}_\mathrm{det}\mathsf{N}^{(2)}_\phi.
\end{align}
In addition, the remaining part involving $\usto\phi_\uuv^{(2)}$ gives rise to \Eq{def Nphi sto}.

\section{Relation between the first and second order UV modes solutions}\label{Sec:B.2}

\subsection{Attractor case}
\label{Sec:B.1}
In the attractor case described by the solution \eqref{solution_SUA_att}, the matrix $\mathbf{B}_\uuv^{(1)}$ defined in \Eq{def B UV} can be rewritten as
\begin{align}
   \mathbf{B}_\uuv^{(1)}(N',k) &= -\frac{\partial {\mathbf{A}}}{\partial \bar{\phi}}\phi_\mathrm{UV}^{(1)}(N',k)-  \frac{\partial {\mathbf{A}}}{\partial \bar{\pi}}\pi_\mathrm{UV}^{(1)}(N',k)\notag\\
   &=-C_a^{(1)}(k)\left(\frac{\partial {\mathbf{A}}}{\partial \bar{\phi}}\bar{\pi}(N')+  \frac{\partial {\mathbf{A}}}{\partial \bar{\pi}}\frac{d\bar{\pi}}{dN'}\right)=-C_a^{(1)}(k)\frac{d {\mathbf{A}}}{dN'}.
    \label{B_phiUV_att}
\end{align}
Substituting \Eq{B_phiUV_att} into the solution \eqref{solution_UV_second} for $\mathbf{\Phi}_{\mathrm{UV}}^{(2)}$, we obtain
\begin{align} \nonumber
   \mathbf{\Phi}_{\mathrm{UV}}^{(2)}(N,k) 
   &=-C_a^{(1)}(k)\int_0^N \dd N' \mathbf{U}(N,N')\frac{d{\mathbf{A}}(N')}{dN'}\mathbf{\Phi}_\mathrm{IR}^{(1)}(N')\notag\\
   &=-C_a^{(1)}(k){\mathbf{A}}(N)\mathbf{\Phi}_\mathrm{IR}^{(1)}(N)\notag\\
   &\quad+C_a^{(1)}(k)\int_0^{N}\dd N'\left(\frac{d\mathbf{U}(N,N')}{dN'}{\mathbf{A}}(N')\mathbf{\Phi}_\mathrm{IR}^{(1)}(N')+\mathbf{U}(N,N'){\mathbf{A}}(N')\frac{d \mathbf{\Phi}_\mathrm{IR}^{(1)}(N')}{dN'}\right)\notag\\
   &=-C_a^{(1)}(k){\mathbf{A}}(N)\mathbf{\Phi}_\mathrm{IR}^{(1)}(N) \notag\\
   &\quad+C_a^{(1)}(k)\int_0^{N}\dd N'\mathbf{U}(N,N'){\mathbf{A}}(N')\left({\mathbf{A}}(N')\mathbf{\Phi}_\mathrm{IR}^{(1)}(N')+\frac{d \mathbf{\Phi}_\mathrm{IR}^{(1)}(N')}{dN'}\right).
   \label{solution_UV_second_att_integrated}
\end{align}
In the second equality, we performed an integration by parts and used the initial conditions \eqref{initial conditions IR}. In the third equality, we used the following identity satisfied by the propagator $\mathbf{U}$:
\begin{align}
    \dif{}{N'}\mathbf{U}(N,N') - \mathbf{U}(N,N'){\mathbf{A}}(N') =0 . \label{equation_inverse_U}
\end{align}
Finally, by using the first-order equations for the IR modes, \eqref{phi IR first}, one obtains the expressions \eqref{phi UV second det ori} and \eqref{phi UV second sto ori}.

\subsection{Non-attractor case}

In this appendix, we extend the attractor analysis of Section \ref{Sec:4} to the full UV-mode solution, including non-attractor contributions, and thereby clarify the structure of the UV modes and the resulting noise terms in the IR system. The discussion proceeds in parallel with the attractor case. The first-order UV-mode solution \eqref{solution_SUA} can be written as
\begin{align}
    \mathbf{\Phi}^{(1)}_\mathrm{UV}(N,k)&=\left(C_a^{(1)}(k)+C_b^{(1)}(k)\int_0^Nf(N')dN'\right)\left[\begin{array}{c}
         \bar{\pi}(N)  \\
       \frac{d\bar{\pi}(N)}{dN}  
    \end{array}\right]+C_b^{(1)}(k)f(N)\left[\begin{array}{c}
         0  \\
       \bar{\pi}(N)  
    \end{array}\right]\notag\\
    &\equiv F^{(1)}(N,k) \left[\begin{array}{c}
         \bar{\pi}(N)  \\
       \frac{d\bar{\pi}(N)}{dN}  
    \end{array}\right]+C_b^{(1)}(k)f(N)\left[\begin{array}{c}
         0  \\
       \bar{\pi}(N)  
    \end{array}\right]
    \label{solution_SUA_non_att}.
\end{align}
Similarly, the second-order UV-mode solution is given by:
\begin{align}
    \mathbf{\Phi}_\mathrm{UV}^{(2)}(N,k) &= \int_0^N \dd N' \mathbf{U}(N,N')\mathbf{B}_\uuv^{(1)}(N',k)\mathbf{\Phi}^{(1)}_\mathrm{IR}(N') \notag \\
    &=-F^{(1)}(N,k){\mathbf{A}}(N)\mathbf{\Phi}_\mathrm{IR}^{(1)}(N) \notag\\
   &\quad+\int_0^{N}\dd N'F^{(1)}(N',k)\mathbf{U}(N,N'){\mathbf{A}}(N')\left({\mathbf{A}}(N')\mathbf{\Phi}_\mathrm{IR}^{(1)}(N')+\frac{d \mathbf{\Phi}_\mathrm{IR}^{(1)}(N')}{dN'}\right)\notag\\
   &\quad+ C_b^{(1)}(k)\int_0^Nf(N')\mathbf{U}(N,N')\left({\mathbf{A}}(N')-\bar{\pi}(N')\frac{\partial {\mathbf{A}}(N')}{\partial \bar{\pi}(N')}\right)\mathbf{\Phi}_\mathrm{IR}^{(1)}(N'),
   \label{solution_UV_second_non_att_integrated}
\end{align}
where we have used that $\mathbf{B}_\uuv^{(1)}(N,k)$ in the non-attractor case is

\begin{align}
   \mathbf{B}_\uuv^{(1)}(N,k) &= -F^{(1)}(N,k)\left(\frac{\partial {\mathbf{A}}}{\partial \bar{\phi}}\bar{\pi}+  \frac{\partial {\mathbf{A}}}{\partial \bar{\pi}}\frac{d\bar{\pi}}{dN}\right) - C_b^{(1)}(k)f(N)\frac{\partial {\mathbf{A}}}{\partial \bar{\pi}}\bar{\pi}\notag\\
   &=-F^{(1)}(N,k)\dif{{\mathbf{A}}}{N}-C_b^{(1)}(k)f(N)\frac{\partial {\mathbf{A}}}{\partial \bar{\pi}}\bar{\pi}.
    \label{B_phiUV_non_att}
\end{align}

As in the attractor case, using the first-order equations for the IR modes, \eqref{phi IR first}, one can decompose the second-order UV-mode solution \eqref{solution_UV_second_non_att_integrated} evaluated at the cutoff scale into a deterministic part and a stochastic part:
\begin{align}
   &\,_\mathrm{det}\mathbf{\Phi}_{\mathrm{UV}}^{(2)}(N,k_\sigma(N)) \notag\\
   &= -F^{(1)}(N,k_\sigma(N)){\widetilde{\mathbf{A}}}(N,k_\sigma(N))\mathbf{\Phi}_\mathrm{IR}^{(1)}(N) \notag\\
   &\quad+ C_b^{(1)}(k_\sigma(N))\int_0^N\dd N'f(N')\mathbf{U}(N,N')\prn{\widetilde{\mathbf{A}}(N',k_\sigma(N)) -\bar{\pi}(N')\frac{\partial \widetilde{\mathbf{A}}(N',k_\sigma(N))}{\partial \bar{\pi}(N')}}\mathbf{\Phi}_\mathrm{IR}^{(1)}(N')\notag\\
   &\quad+ C_b^{(1)}(k_\sigma(N))\int_0^N\dd N'f(N')\mathbf{U}(N,N')\com{\begin{array}{cc}
       0 &  0\\
       0 & 1
   \end{array}}\mathbf{\Xi}^{(1)}(N'), \label{phi UV second det nonattractor}
\end{align}
\begin{align}
   \,_\mathrm{sto}\mathbf{\Phi}_{\mathrm{UV}}^{(2)}(N,k_\sigma(N))
   &= \int_0^{N}\dd N'F^{(1)}(N',k_\sigma(N))\mathbf{U}(N,N'){\mathbf{A}}(N',k_\sigma(N))\mathbf{\Xi}^{(1)}(N')\notag\\
   &\quad- C_b^{(1)}(k_\sigma(N))\int_0^N\dd N'f(N')\mathbf{U}(N,N')\com{\begin{array}{cc}
       0 &  0\\
       0 & 1
   \end{array}}\mathbf{\Xi}^{(1)}(N')\notag\\
   &\quad+   \int_0^{N}\dd N'F^{(1)}(N',k_\sigma(N))\mathbf{U}(N,N'){\mathbf{A}}(N',k_\sigma(N)){\mathbf{W}}(N',k_\sigma(N))\mathbf{\Phi}_\mathrm{IR}^{(1)}(N')\notag\\
   &\quad+ C_b^{(1)}(k_\sigma(N))\int_0^Nf(N')\mathbf{U}(N,N')\mathbf{W}(N',k_\sigma(N))\mathbf{\Phi}_\mathrm{IR}^{(1)}(N').\label{phi UV second sto nonattractor}
\end{align}

To see that \Eq{phi UV second det nonattractor} is indeed equivalent to the definition of the deterministic part in terms of functional derivatives, \eqref{def UV second det}, we will use the first-order UV-mode solution \eqref{solution_SUA_non_att} into the definition of $\udet\mathbf{\Phi}^{(2)}_\uuv(N,k_\sigma(N))$ in terms of functional derivatives \eqref{def UV second det}. After some straightforward computation, one can show that $\udet\mathbf{\Phi}^{(2)}_\uuv(N,k_\sigma(N))$ follows the following differential equation at leading order in gradient expansion

\begin{align}
    &\prn{\pdif{}{N}  + \widetilde{\mathbf{A}}(N)} \udet\mathbf{\Phi}^{(2)}_\uuv(N,k_\sigma(N)) \notag\\
    &= -\prn{\pdif{\widetilde{\mathbf{A}}}{\bar\phi}\phi_\uir^{(1)} + \pdif{\widetilde{\mathbf{A}}}{\bar\pi}\pi^{(1)}_\uir} \mathbf{\Phi}_\uuv^{(1)}(N,k_\sigma(N))\notag\\
    &\quad- F^{(1)}(N,k_\sigma(N))\widetilde{\mathbf{A}}(N)\mathbf{\Xi}^{(1)}(N) + C^{(1)}_b(k_\sigma(N))f(N)\com{\begin{array}{cc}
        0 &  0\\
        0 & 1
    \end{array}}\mathbf{\Xi}^{(1)}(N)\notag\\
    &= - F^{(1)}(N,k_\sigma(N))\dif{\widetilde{\mathbf{A}}(N)}{N}\mathbf{\Phi}_\uir^{(1)} (N)- C^{(1)}_b(k_\sigma(N))f(N)\bar\pi(N) \pdif{\widetilde{\mathbf{A}}(N)}{\bar\pi}\mathbf{\Phi}_\uir^{(1)}(N)\notag\\
    &\hspace{20mm} - F^{(1)}(N,k_\sigma(N))\widetilde{\mathbf{A}}(N)\mathbf{\Xi}^{(1)}(N) + C^{(1)}_b(k_\sigma(N))f(N)\com{\begin{array}{cc}
        0 & 0\\
        0 & 1
    \end{array}}\mathbf{\Xi}^{(1)}(N).
    \label{functional deri phiUV 2 eq}
\end{align}
Here we have used the first-order IR equation \eqref{IR modes 1st}, and noted that $f(N)$ is defined by \Eq{def_F} so that \Eq{solution_SUA_non_att} satisfies \Eq{UV modes 1st}. We also employ the following relations among
the components of $\tilde{\mathbf{A}}$: $\widetilde{A}_{11} = \partial_{\bar{\phi}} \widetilde{A}_{12} = \partial_{\bar{\pi}} \widetilde{A}_{12} = 0$, and $\partial_{\bar{\pi}} \widetilde{A}_{21}=\partial_{\bar{\phi}} \widetilde{A}_{22}$ under the separate-universe assumption, together with the commutativity of partial derivatives. In addition, the equation: $\bar\pi\dif{f}{N} + f\prn{2\dif{\bar\pi}{N} - \widetilde{A}_{22}\bar\pi   }=0$ is used, which can be derived by substituting \eqref{solution_SUA_non_att} into \Eq{UV modes 1st}.

The formal solution of \Eq{functional deri phiUV 2 eq}, with vanishing initial conditions, is given by
\begin{align}
    &\udet\mathbf{\Phi}^{(2)}_\uuv(N,k_\sigma(N)) \notag\\
    &= -\int_0^N \dd N' \mathbf{U}(N,N')\left(F^{(1)}(N',k_\sigma(N))\dif{\widetilde{\mathbf{A}}}{N'}\mathbf{\Phi}_\uir^{(1)}(N') +C^{(1)}_b(k_\sigma(N))f(N')\bar\pi(N') \pdif{\widetilde{\mathbf{A}}(N')}{\bar\pi}\mathbf{\Phi}_\uir^{(1)}(N')\right.\notag\\
    &\hspace{20mm} \left. - F^{(1)}(N',k_\sigma(N))\widetilde{\mathbf{A}}(N')\mathbf{\Xi}^{(1)}(N') + C^{(1)}_b(k_\sigma(N))f(N)\com{\begin{array}{cc}
        0 & 0\\
        0 & 1
    \end{array}}\mathbf{\Xi}^{(1)}(N)\right).
\end{align}
By integrating the first term by parts and using \Eq{equation_inverse_U}, one can show that this expression identical to \Eq{phi UV second det nonattractor}.

\bibliographystyle{JHEP}
\bibliography{ref}
\end{document}